# Queues Are Databases

Jim Gray

December 1995

Technical Report

MSR-TR-95-56

Microsoft Research

Microsoft Corporation

One Microsoft Way

Redmond, WA  98052

# Queues are Databases[1]

Jim Gray

Microsoft Research, 301 Howard St., S.F., CA. 94105

Gray@Microsoft.com

December, 1995

## Abstract:

Message-oriented-middleware (MOM) has become an small industry.  MOM offers queued transaction processing as an advance over pure client-server transaction processing.  This note makes four points:

**Queued transaction processing is less general than direct transaction processing.** Queued systems are built on top of direct systems.  You cannot build a direct system atop a queued system.  It is difficult to build direct, conversational, or distributed transactions atop a queued system.

**Queues are *interesting* databases with *interesting* concurrency control**.  It is best to build these mechanisms into a standard database system so other applications can use these interesting features.

**Queue systems need DBMS functionality**.  Queues need security, configuration, performance monitoring, recovery, and reorganization utilities.   Database systems already have these features.  A full-function MOM system duplicates these database features.

**Queue managers are simple TP-monitors managing server pools driven by queues**. Database systems are encompassing many server pool features as they evolve to TP-lite systems.

---

[1] **Acknowledgments**: These ideas derive from discussions with Andrea Borr (Oracle), Richard Carr (Tandem), Dieter Gawlick (Oracle), Pat Helland (Microsoft), Franco Putzolu (Oracle),  Andreas Reuter (U. Stuttgart) and Bill Highliman (NetWeave).

## 1. Queues Are Best Built Atop Direct TP Systems.

TP systems offer both *queued* and *direct* transaction processing. They offer *both client-server* and *peer-to peer* direct processing. Gray & Reuter [pp. 246] offer the following common taxonomy of process-to-process communication:

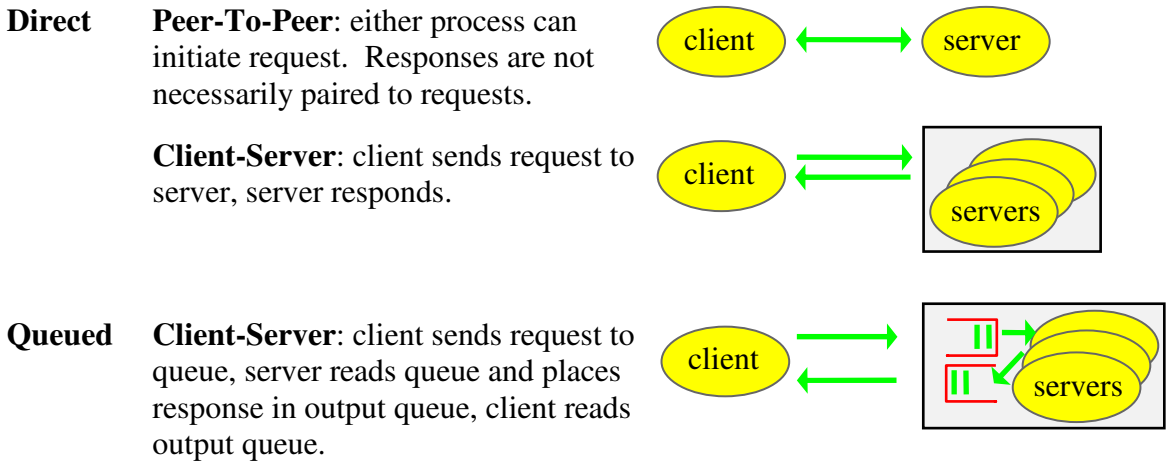

**Direct**  **Peer-To-Peer**: either process can initiate request. Responses are not necessarily paired to requests.

**Client-Server**: client sends request to server, server responds.

**Queued**  **Client-Server**: client sends request to queue, server reads queue and places response in output queue, client reads output queue.

The shaded boxes in the figure represent a dispatcher that binds client requests to servers. The dispatcher was traditionally called a TPmonitor. Today dispatchers are called Object-Request Brokers (ORBs).

In queued processing, clients place request messages in a queue. A pool of server processes, managed by a TP-monitor, service these request queues, perhaps placing results in other queues. Clients can poll these output queues to see the status or outcome of their transaction requests. If the client and server on different computers, the queue may be replicated at both the client and server node so that either end can generate and process messages even if when disconnected.

Queued processing is the basic mechanism of IBM's IMS, so we have 30 years experience with its pros and cons. Advocates of queued processing point out that, at saturation, a direct system is really a queued system: the TPmonitor dispatches servers via a queuing mechanism to do load control. When clients saturate a server pool, the queues become visible. Indeed, it is optimal to schedule new requests to a server pool via a single global queue.

The difficulty is that queued transaction processing of a request-response is three ACID units:

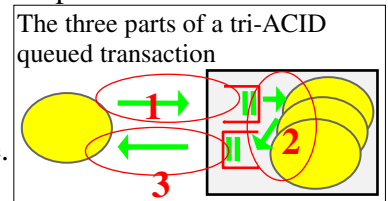
The three parts of a tri-ACID queued transaction

1. Client places request in queue.
2. Server dequeues request, performs task, enqueues response.
3. Requester dequeues response from output queue .

This tri-ACID unit model has the benefit of decoupling the client from the server, but has the liability that it makes multi-request (conversational) transactions impossible. Since each message exchange is three ACID units, one cannot wrap a multi-step dialog in a single ACID transaction unit. Implementing distributed transactions, conversational transactions, or multi-step transactions on top of a queued system requires building a lot of application-level machinery. IMS customers have invested millions of dollars in such efforts.

By contrast, direct transaction processing systems can easily add a queuing mechanism as a new transaction type. They implement a direct transaction that places requests in queues, have pools of servers that poll these queues, and have a third transaction that queries the output queues. This is the approach that CICS, ACMS, Tuxedo, Topend, and Encina take. At last count, CICS had over six distinct queue managers as part of the regular product -- each with slightly different performance-functionality tradeoffs.

I am not arguing that queued processing is bad -- quite the contrary. Queued processing has been a common transaction processing style and will continue to be very important in the future. Queued processing is increasingly important for workflow, disconnected operation, and replication applications. It has always been the mainstay of batch and spool operations. Using Bill Highliman's acronyms[2], the world needs both MOM and DAD.

---

[2] DAD: direct access to data, MOM: message oriented middleware.

HPTS 95 Position Paper:  Queues are Databases                                                                                   4

The controversial opinion is that a queue manager is best built as a naive resource manager atop an object-relational database system [Chamberlin], [Stonebraker]. That system must have good concurrency control, recovery, triggers, security, operations interfaces, and utilities -- indeed it must be a good TP-lite system. Given such a base, a queue manager would be one of the first class libraries I would write.

## 2. Queues Are "Interesting" Databases

Storing queues in a database has considerable appeal. The idea is that queues are a database class encapsulated with create(), enqueue(), dequeue(), poll(), and destroy() methods. By using a database, the queue manager becomes a naive resource manger with no special code for startup, shutdown, checkpoint, commit, query, security, or utilities. Rather it is just a simple application – the database system does all the hard stuff like locking, logging, access paths, recovery, utilities, security, performance monitoring, and so on. The queue manager benefits from all the database utilities to query, backup, restore, reorganize, and replicate data. In addition it piggybacks on the TP-lite and trigger mechanisms of the database system for process and server pool management.

Queues pose difficult problems when implemented atop a database.

**Performance:** An enqueue transaction is an insert followed by a commit. This places extreme performance demands on the concurrency control and recovery components of a database -- it exposes hotspots and high-overhead code.

**Concurrency control:** The dequeue transaction typically involves deleting a record from the queue, processing the request, enqueuing results in other queues, and then committing. Serializable isolation requires that there can be at most one dequeue executing at a time against each queue. This suggests that queues need lower, indeed specialized, isolation levels.

Gray and Reuter [ibid. pp. 402] outline the concurrency control mechanisms needed to implement queues within a database:

**Read_Past** locks allow a program to skip over dirty (uncommitted records) to find the first committed record. This is what a dequeue() operation wants.

**Read_Through** locks allow a program to examine records that have not yet been committed. This is useful in polling the status of a queued request that is currently being processed.

**Notify** (events) allow a program to wait for a state change in a lock. This allows a dequeue() operation to wait for one or more queues to become non-empty.

Non-transactional queues are sometimes needed for performance reasons. The same reasons encourage us to support non-transactional tables in an SQL database. These tables and queues are not durable (do not survive system restart or media failure), but have low overhead.

The paradox is that queues are just an application data structure. Their concurrency control and recovery needs appear in many other contexts. An auction application looking for a set of sellers to match a buyer needs exactly these features. An emergency dispatch application needs to find the highest-priority request not yet being serviced. Similar requirements appear in workflow, CASE, and parallel programming models like Linda.

There is a pattern here. Each new requirement for a queuing system seems to reflect a corresponding requirement for user-application data. This recurs when one considers, query interfaces to queues, queue performance monitoring, queue backup, restore, recovery utilities, queue security, and so on. Indeed, Richard Carr reports that when a queuing mechanisms was added to Tandem's database servers, several applications became simpler and faster.

## 3. Queue Managers Are Simple TP-Monitors

So far, the discussion has ignored the question of server pool management (threads or processes). Some queues have a server pool attached to them. The servers in this pool are dedicated to servicing entries in the queue. You see this in batch job schedulers, print spoolers, and in many transaction processing systems. TP-monitors (aka: ORBs) configure, manage, and load-balance these pools.

Server pools are configured with a minimum and maximum number of servers. The pool starts at its minimum size. As traffic grows, the pool grows. As traffic shrinks, the pool shrinks. If a server fails, a new server is allocated. If too many servers fail in a time window, the TP-monitor declares the queue broken and human intervention is required. Operator and programmatic interfaces are defined to create, configure, query, and control (start, stop, redefine) queues. This is what the "gray boxes" are doing in Figure 1.

Queued processing has many variants:
 **Periodic**: Servers are created at certain times.
 **Event**: Servers are created on demand when a request first arrives in a queue.
 **Batch**: Servers are created when the queued grows to a certain size.

The queue scheduling policy is often a priority scheme whereby some queue elements are processed before others.

Gee! This sounds like a lot of stuff you do not find in your database system: server pools, timers, priority,….

But, what about triggers and stored procedures? All modern database systems allow users to associate procedures with data records. These procedures fire when the client explicitly invokes the procedure. Triggers implicitly fire when records are read, inserted, deleted, or updated. Stored procedures may be synchronous or asynchronous. Triggers fire at the time of the operation (immediate), or at the time of commit (deferred). They may execute within the ACID transaction of the operation that fired the trigger, or they may begin a new top-level asynchronous transaction.

Stored and trigger procedures are out-calls from the DBMS. They are written in C, COBOL, FORTRAN, Visual Basic, or the DBMS procedural language (Transact SQL or PL/SQL or …). User-written procedures are optionally executed in a protection domain separate from the requester and separate from the DBMS. They are typically executed in a separate process (address space). Sybase's OpenServer design is typical of this idea -- although it uses a single multi-threaded process rather than having a separate protection domain per trigger. Oracle's Rdb uses a separate process per user to manage outcalls.

Managing trigger processes is a chore. For good performance, they must be pre-allocated. There must be load-control to prevent saturation. The pools must grow and shrink with demand. Gradually, the trigger-execution mechanism of the DBMS merges with the DBMS's TP-lite front-end dispatcher to make a fairly general TP-lite monitor. Indeed, the Sybase OpenServer started as a front-end, then became a side-end (trigger) and back-end (gateway) mechanism.

So DBMS systems are growing a server pool management system. This is part of the evolution of DBMS to TP-lite to TP-heavy.

Not much is needed to add queued processing to a TP-lite database system. First one must implement queues as an encapsulated type atop the Object-Relational system. Then one must recognize that triggers may be fired as part of a transaction, or fired asynchronously as a new ACID unit (either immediately, or if and when the transaction commits). This gives a simple queued transaction processing system. If it solves the concurrency and performance problems, should be as scaleable and robust as the underlying DBMS.

## 4. Summary

Many people are building queue managers on file systems as a transactional resource manager and a TP-lite monitor.  An alternative approach is to evolve an Object-Relational database system to support the mechanisms needed to build a queuing system:

- reduced isolation levels and fine granularity locking.
- efficient support for simple transactions.
- asynchronous trigger invocation executed by server pools.
- server pools management and dispatching.

These basic facilities enable the implementation of queue managers but also make the DBMS more useful to other applications.

## 5. Reconsideration

This position paper was intended to generate controversy at the High Performance Transaction Processing Workshop (HPTS).   Amazingly, everyone either agreed or was so disgusted that they left the room.   In the end, there was no heated discussion.   I was astonished.   In defense of MOM, one must point out that my discussion assumes a homogeneous environment: one ubiquitous database and transport.  In essence it says: if you got universal DAD you can build MOM.

The flaw in this argument seems to be that DAD presumes that the database system has direct access from everywhere to everywhere and has storage everywhere.  The database system has to have a presence on Prime, Apollo, Unisys, Boroughs, NCR, Singer, Harris, Tandem, Prime, Apollo, Sequent, Next, NetWare, DOS,… Queues need to be stored on both client and server.  Thus, the DBMS must exist on *all* these exotic platforms (no offense intended).  Many of the MOM companies make their living by porting a minimalist database (a queue system) to these exotic platforms for a fee.  The cost of porting a full-blown DBMS to such systems is prohibitive.

My conclusion from this are:

1. DADs (OR-DBMSs) will evolve to provide queues. Portable systems like Oracle, Sybase, Informix, DB2/CS will offer heterogeneous queuing among commodity platforms.
2. MOMs will thrive by connecting exotic heterogeneous systems together.